\documentclass[12pt,a4paper]{article}
\usepackage{amssymb,amsmath}
\usepackage[dvips]{lscape,graphicx}

\newcommand{\bc}{\begin{center}}
\newcommand{\ec}{\end{center}}
\newcommand{\bd}{\begin{displaymath}}
\newcommand{\ed}{\end{displaymath}}
\newcommand{\be}{\begin{equation}}
\newcommand{\ee}{\end{equation}}
\newcommand{\ba}{\begin{array}}
\newcommand{\ea}{\end{array}}
\newcommand{\bt}{\begin{tabular}}
\newcommand{\et}{\end{tabular}}

\newcommand{\ds}{\displaystyle}

\begin{document}

\title{Quasi--fixed point scenarios and the Higgs mass in the $E_6$ inspired SUSY models}

\author{R.~Nevzorov${}^{a,\,b}$\\[1mm]
\itshape{$^a$ Theory Department,  ITEP, Moscow, 117218, Russia}\\[2mm]
\itshape{$^b$ ARC Centre of Excellence for Particle Physics at the Terascale and CSSM,}\\[0mm]
\itshape{School of Chemistry and Physics, The University of Adelaide,}\\[0mm]
\itshape{Adelaide, South Australia 5005, Australia}}

\date{}

\maketitle

\begin{abstract}{
\noindent
We analyse the two--loop renormalization group (RG) flow of the gauge and Yukawa couplings within the $E_6$ inspired
supersymmetric (SUSY) models with extra $U(1)_{N}$ gauge symmetry under which right--handed neutrinos
have zero charge. In these models single discrete $\tilde{Z}^{H}_2$ symmetry forbids the tree-level
flavor-changing transitions and the most dangerous baryon and lepton number violating operators.
We consider two different scenarios A and B that involve extra matter beyond the MSSM contained in
three and four $5+\overline{5}$ representations of $SU(5)$ respectively plus three $SU(5)$ singlets which carry
$U(1)_{N}$ charges. In the scenario A the measured values of the $SU(2)_W$ and $U(1)_Y$ gauge couplings lie
near the fixed points of the RG equations. In the scenario B the contribution of two--loop
corrections spoils the unification of gauge couplings resulting in the appearance of the Landau pole
below the Grand Unification scale $M_X$. The solutions for the Yukawa couplings also approach the quasi fixed
points with increasing their values at the scale $M_X$. We calculate the two--loop upper bounds on the lightest
Higgs boson mass in the vicinity of these quasi fixed points and compare the results of our analysis with the
corresponding ones in the NMSSM. In all these cases the theoretical restrictions on the SM--like Higgs boson
mass are rather close to $125\,\mbox{GeV}$.
}
\end{abstract}


\newpage

\section{Introduction}

\noindent
The recent discovery of a SM-like Higgs state with a mass around $\sim 125$~GeV~
\cite{:2012gk,:2012gu} is consistent with the supersymmetric (SUSY) extensions of the
Standard Model (SM). Indeed, in the minimal supersymmetric standard model (MSSM)
the mass of the lightest Higgs particle, which manifests itself in the interactions
with gauge bosons and fermions as a SM--like Higgs boson, does not exceed
$130-135\,\mbox{GeV}$. Although the MSSM is one of the most attractive and best
studied extensions of the SM it suffers from the $\mu$ problem: the superpotential
of the MSSM contains one bilinear term $\mu H_d H_u$ which is present
before SUSY is broken. Thus one would naturally expect the parameter $\mu$ to be
of the order of the Planck scale $M_{\rm{Pl}}$. On the other hand in order to get
the correct pattern of electroweak (EW) symmetry breaking (EWSB), $\mu$ is required
to be of the order of the EW scale.

An elegant solution of the $\mu$ problem naturally arises in the framework of $E_6$
inspired models. At high energies $E_6$ can be broken down to the rank-5 gauge group
that leads to low energy gauge symmetry with additional $U(1)'$ factor in comparison
to the SM. The remaining $U(1)'$ symmetry is a linear combination of $U(1)_{\psi}$
and $U(1)_{\chi}$
\be
U(1)'=U(1)_{\chi}\cos\theta+U(1)_{\psi}\sin\theta\,.
\label{1}
\ee
Two anomaly-free $U(1)_{\psi}$ and $U(1)_{\chi}$ symmetries are defined by:
$E_6\to SO(10)\times U(1)_{\psi}$, $SO(10)\to SU(5)\times U(1)_{\chi}$.
If $\theta\ne 0\,\mbox{or}\,\,\pi$ the extra $U(1)'$ gauge symmetry forbids an elementary
$\mu$ term but allows an interaction of the extra SM singlet superfield $S$ with
the Higgs doublet supermultiplets $H_d$ and $H_u$ in the superpotential:
$\lambda S H_d H_u$. At the TeV scale the scalar component of the SM
singlet superfield $S$ acquires a non-zero vacuum expectation value (VEV) breaking
$U(1)'$ and an effective $\mu$--term of the required size is automatically generated.

Here we focus on the supersymmetric extension of the SM which is based on the low--energy
SM gauge group together with an extra $U(1)_{N}$ gauge symmetry that corresponds to the
angle $\theta=\arctan\sqrt{15}$ in Eq.~(\ref{1}). Only in this Exceptional Supersymmetric
Standard Model (E$_6$SSM) \cite{King:2005jy}--\cite{King:2005my} right--handed neutrinos
do not participate in the gauge interactions. As a consequence they may be superheavy,
shedding light on the origin of the mass hierarchy in the lepton sector. Because right--handed
neutrinos are allowed to have large masses, they may decay into final states with lepton number
$L=\pm 1$, thereby creating a lepton asymmetry in the early Universe that subsequently gets
converted into the observed baryon asymmetry through the EW phase transition \cite{King:2008qb}.

To ensure that E$_6$SSM is anomaly--free, the particle spectrum in this extension of the SM
is extended to fill out three complete 27-dimensional representations of the gauge group $E_6$.
Each $27$-plet contains one generation of ordinary matter; singlet fields, $S_i$; up
and down type Higgs doublets, $H^{u}_{i}$ and $H^{d}_{i}$; charged $\pm 1/3$ coloured
exotics $D_i$, $\bar{D}_i$. The matter content and correctly normalized Abelian charge assignment
are summarised in Table~\ref{charges}. To suppress tree-level flavor-changing transitions and the
most dangerous baryon and lepton number violating operators in the E$_6$SSM an approximate
$Z^{H}_2$ symmetry can be imposed. Under this symmetry all superfields except one pair of $H^{u}_{i}$
and $H^{d}_{i}$ (i.e. $H_{u}$ and $H_{d}$) and one of the SM-type singlet superfields $S_i$ (i.e. $S$)
are odd.  When all $Z^{H}_2$ symmetry violating couplings are small this discrete symmetry allows
to suppress flavour changing processes. If the Lagrangian of the E$_6$SSM is invariant with respect
to either a $Z_2^L$ symmetry, under which all superfields except leptons are even, or a $Z_2^B$ discrete
symmetry that implies that exotic quark and lepton superfields are odd whereas the others remain even,
then the most dangerous baryon and lepton number violating operators get forbidden and proton is
sufficiently longlived \cite{King:2005jy}--\cite{King:2005my}. The presence of exotic matter predicted
by the E$_6$SSM at the TeV scale may lead to spectacular new physics signals at the LHC which
were analysed in \cite{King:2005jy}--\cite{King:2005my}, \cite{Accomando:2006ga}. Recently the particle
spectrum and collider signatures associated with it were studied within the constrained version of
the E$_6$SSM (cE$_6$SSM) \cite{cE6SSM}. The threshold corrections to the running gauge and Yukawa
couplings in the E$_6$SSM and cE$_6$SSM were studied in details in \cite{Athron:2012pw}.
The renormalization of VEVs in the E$_6$SSM was considered in \cite{Sperling:2013eva}.

\begin{table}[ht]
  \centering
  \begin{tabular}{|c|c|c|c|c|c|c|c|c|c|c|c|}
    \hline
 & $Q$ & $u^c$ & $d^c$ & $L$ & $e^c$ & $N^c$ & $S$ & $H^u$ & $H^d$ & $D$ &
 $\overline{D}$  \\
 \hline
$\sqrt{\frac{5}{3}}Q^{Y}_i$
 & $\frac{1}{6}$ & $-\frac{2}{3}$ & $\frac{1}{3}$ & $-\frac{1}{2}$
& $1$ & $0$ & $0$ & $\frac{1}{2}$ & $-\frac{1}{2}$ & $-\frac{1}{3}$ &
 $\frac{1}{3}$  \\
 \hline
$\sqrt{{40}}Q^{N}_i$
 & $1$ & $1$ & $2$ & $2$ & $1$ & $0$ & $5$ & $-2$ & $-3$ & $-2$ &
 $-3$  \\
 \hline
  \end{tabular}
  \caption{\it\small The $U(1)_Y$ and $U(1)_{N}$ charges of matter fields in the
   $E_6$ inspired SUSY models with extra $U(1)_{N}$ gauge symmetry.}
  \label{charges}
\end{table}

In this article we explore the two--loop renormalisation group (RG) flow of the gauge and Yukawa couplings
within the $E_6$ inspired supersymmetric extensions of the SM with extra $U(1)_{N}$ gauge symmetry
in which a single discrete $\tilde{Z}^{H}_2$ symmetry forbids tree-level flavor-changing transitions and the
most dangerous baryon and lepton number violating operators \cite{nevzorov}. Two different scenarios A and B,
that involve extra matter beyond the MSSM contained in three and four $5+\overline{5}$ representations of
$SU(5)$ respectively together with three SM singlets with $U(1)_{N}$ charges, are considered. These scenarios
lead to different phenomenological implications associated with the exotic quarks $D_i$ and $\bar{D}_i$.
In the case of scenario A we demonstrate that the solutions of the RG equations for the $SU(2)_W$ and $U(1)_Y$
gauge couplings tend to converge towards the quasi--fixed points which are rather close to the experimentally
measured low energy values of these couplings while the convergence of the corresponding solutions for the
strong gauge coupling to the quasi--fixed point is rather weak. In the scenario B the values of the strong gauge
coupling $g_3(Q)$ near the EW scale tend to be substantially smaller than the experimentally measured central
value of this coupling. This implies that the values of $\alpha_3(M_Z)$, which are within one standard deviation of
its measured central value, result in the appearance of the Landau pole below the GUT scale in this scenario.
Thus the gauge coupling unification gets basically spoiled by large two--loop corrections in this case.

We also argue that the solutions for the Yukawa couplings approach the quasi--fixed points with increasing
their values at the Grand Unification (GUT) scale $M_X$. In contrast with the MSSM the quasi--fixed point scenarios
in the SUSY models being considered here, that correspond to $\tan\beta\sim 1$, have not been ruled out.
In other words these scenarios can lead to the solutions with the SM-like Higgs mass around $\sim 125\,\mbox{GeV}$.
We calculate the two--loop upper bounds on the lightest Higgs boson mass in the vicinity of the quasi--fixed points
in these models and compare the obtained results with the corresponding ones in the NMSSM.
Although we focus primarily on the part of the parameter space, where the lightest Higgs boson mass attains its
maximal value in the SUSY models mentioned above (see, for example \cite{Ellwanger:2006rm}--\cite{Masip:1998jc}),
our analysis indicates that the values of the Yukawa couplings near the quasi--fixed points are such that the
SM--like Higgs state has a mass which is lower than $130\,\mbox{GeV}$ for TeV stop masses.

The layout of the remainder of the paper is as follows. In the next section we briefly review
the $E_6$ inspired SUSY models with exact custodial $\tilde{Z}^{H}_2$ symmetry. In section 3 the RG flow
of the gauge and Yukawa couplings is studied and the two--loop upper bounds on the lightest Higgs boson mass
in the vicinity of the quasi--fixed points are calculated. Section 4 concludes the paper.

\section{$E_6$ inspired SUSY models with exact $\tilde{Z}^{H}_2$ symmetry}

In this section, we briefly review the $E_6$ inspired SUSY models with exact custodial $\tilde{Z}^{H}_2$
symmetry \cite{nevzorov}. These models imply that near some high energy scale ($M_X$) $E_6$ or
its subgroup is broken down to $SU(3)_C\times SU(2)_W\times U(1)_Y\times U(1)_{\psi}\times U(1)_{\chi}$.
Below GUT scale $M_X$ the particle content of the considered models involves three copies of $27_i$--plets
and a set of $M_{l}$ and $\overline{M}_l$ supermultiplets from the incomplete $27'_l$ and $\overline{27'}_l$
representations of $E_6$\footnote{Because multiplets $M_{l}$ and $\overline{M}_l$ have opposite $U(1)_{Y}$,
$U(1)_{\psi}$ and $U(1)_{\chi}$ charges their contributions to the anomalies get cancelled identically.}.
All matter superfields, that fill in complete $27_i$--plets, are odd under $\tilde{Z}^{H}_2$ discrete symmetry
while the supermultiplets $\overline{M}_l$ can be either odd or even. All supermultiplets $M_{l}$ are even
under the $\tilde{Z}^{H}_2$ symmetry and therefore can be used for the breakdown of gauge symmetry.
In order to ensure that the $SU(2)_W\times U(1)_Y\times U(1)_{\psi}\times U(1)_{\chi}$ symmetry is broken
down to $U(1)_{em}$ associated with the electromagnetism the set of multiplets $M_{l}$ should involve
$H_u$, $H_d$, $S$ and $N^c_H$.

These $E_6$ inspired SUSY models also imply that just below the GUT scale $U(1)_{\psi}\times U(1)_{\chi}$
gauge symmetry is broken by the VEVs of $N^c_H$ and $\overline{N}_H^c$ down to $U(1)_{N}\times Z_{2}^{M}$, where
$Z_{2}^{M}=(-1)^{3(B-L)}$ is a matter parity. This can be easily arranged because matter parity is a discrete
subgroup of $U(1)_{\psi}$ and $U(1)_{\chi}$. Such breakdown of $U(1)_{\psi}$ and $U(1)_{\chi}$ gauge symmetries
guarantees that the exotic states which originate from $27_i$ representations of $E_6$ as well as ordinary quark
and lepton states survive to low energies. The large VEVs of $N^c_H$ and $\overline{N}_H^c$ can induce the large
Majorana masses for right-handed neutrinos allowing them to be used for the see--saw mechanism. Since $N^c_H$ and
$\overline{N}_H^c$ acquire VEVs both supermultiplets must be even under the $\tilde{Z}^{H}_2$ symmetry.

Here we restrict our consideration to the simplest scenarios in which $\overline{H}_u$, $\overline{H}_d$ and
$\overline{S}$ are odd under the $\tilde{Z}^{H}_2$ symmetry and $\overline{S}$ from the $\overline{27'}_l$ gets
combined with the superposition of the corresponding components from $27_i$ resulting in the vectorlike states
with masses of order of $M_X$. At low energies (i.e. TeV scale) the superfields $H_u$, $H_d$ and $S$ play the role
of Higgs fields. The VEVs of these superfields ($\langle H_d \rangle = v_1/\sqrt{2}$, $\langle H_u \rangle = v_2/\sqrt{2}$
and $\langle S \rangle = s/\sqrt{2}$) break the $SU(2)_W\times U(1)_Y\times U(1)_{N}$ gauge symmetry down to
$U(1)_{em}$. The $\tilde{Z}^{H}_2$ symmetry allows the Yukawa interactions in the superpotential that originate from
$27'_l \times 27'_m \times 27'_n$ and $27'_l \times 27_i \times 27_k$. Since the set of multiplets $M_{l}$ contains
only one pair of doublets $H_d$ and $H_u$ the $\tilde{Z}^{H}_2$ symmetry defined above forbids flavor-changing processes
at the tree level. Nonetheless if the set of $\tilde{Z}^{H}_2$ even supermultiplets $M_{l}$ involve only $H_u$, $H_d$,
$S$ and $N^c_H$ then the lightest exotic quarks are extremely long--lived particles because $\tilde{Z}^{H}_2$ symmetry
forbids all Yukawa interactions in the superpotential that can allow the lightest exotic quarks to decay\footnote{The
models with stable charged exotic particles are ruled out by different terrestrial experiments \cite{42}.}.

To ensure that the lightest exotic quarks decay within a reasonable time the set of $\tilde{Z}^{H}_2$ even supermultiplets
$M_{l}$ can be supplemented by either $L_4$ (scenario A) or $d^c_4$ (scenario B). In both cases it is assumed that at low
energies extra matter beyond the MSSM fill in complete $SU(5)$ representations to preserve gauge coupling unification
which remains almost exact in the one--loop approximation if this condition is fulfilled. In the scenario A this requires
that $\overline{H}_u$ and $\overline{H}_d$ from the $\overline{27'}_l$ get combined with the superposition of the
corresponding components from $27_i$ forming vectorlike states which gain masses $\sim M_X$. The supermultiplets
$L_4$ and $\overline{L}_4$ are also expected to form vectorlike states. However these states are required to be light enough
to ensure that the lightest exotic quarks decay sufficiently fast\footnote{The appropriate mass term $\mu_L L_4\overline{L}_4$
in the superpotential can be induced within SUGRA models just after the breakdown of local SUSY if the K\"ahler potential
contains an extra term $(Z_L (L_4\overline{L}_4)+h.c)$\cite{45}.}. In this case the baryon and lepton number conservation
requires exotic quarks to be leptoquarks. The low energy matter content in the scenario A involves:
\be
\ba{c}
3\left[(Q_i,\,u^c_i,\,d^c_i,\,L_i,\,e^c_i)\right]
+3(D_i,\,\bar{D}_i)+2(S_{\alpha})+2(H^u_{\alpha})+2(H^d_{\alpha})\\[2mm]
+L_4+\overline{L}_4+S+H_u+H_d\,,
\ea
\label{2}
\ee
where $\alpha=1,2$ and $i=1,2,3$. Neglecting all suppressed non-renormalisable interactions one gets an explicit expression
for the superpotential in this case
\be
\ba{c}
W_{A} = \lambda S (H_u H_d) + \lambda_{\alpha\beta} S (H^d_{\alpha} H^u_{\beta})
+ \kappa_{ij} S (D_{i} \overline{D}_{j}) + \tilde{f}_{\alpha\beta} S_{\alpha} (H^d_{\beta} H_u)
+ f_{\alpha\beta} S_{\alpha} (H_d H^u_{\beta}) \\[2mm]
+ g^D_{ij} (Q_i L_4) \overline{D}_j
+ h^E_{i\alpha} e^c_{i} (H^d_{\alpha} L_4) + \mu_L L_4\overline{L}_4
+ W_{MSSM}(\mu=0)\,.
\ea
\label{3}
\ee

In the scenario B extra matter beyond the MSSM fill in complete $SU(5)$ representations if
$\overline{H}_u$, $\overline{H}_d$, $d^c_4$ and $\overline{d^c}_4$ survive to the TeV scale.
In the simplest case $\overline{H}_u$ and $\overline{H}_d$ are odd under the $\tilde{Z}^{H}_2$
symmetry so that they do not acquire VEVs. In contrast, $d^c_4$ and $\overline{d^c}_4$ are
expected to be $\tilde{Z}^{H}_2$ even superfields since these supermultiplets should give rise
to the decays of the lightest exotic color states. In this case the exotic quarks are allowed
to have non-zero Yukawa couplings with pair of quarks. They can also interact with $d^c_4$ and
right-handed neutrinos. If Majorana right-handed neutrinos are very heavy ($\sim M_X$) then
the interactions of exotic quarks with leptons are extremely suppressed so that $\overline{D}_i$
and $D_i$ manifest themselves in the Yukawa interactions as superfields with baryon number
$\left(\pm\dfrac{2}{3}\right)$. When the Yukawa couplings of $d^c_4$ are small enough (i.e. less
than $10^{-5}-10^{-4}$) then the baryon and lepton number violating operators are suppressed
and proton is sufficiently longlived. In the scenario B the low energy matter content may be
summarized as:
\be
\ba{c}
3\left[(Q_i,\,u^c_i,\,d^c_i,\,L_i,\,e^c_i)\right]
+3(D_i,\,\bar{D}_i)+3(H^u_{i})+3(H^d_{i})+2(S_{\alpha})\\[2mm]
+d^c_4+\overline{d^c}_4+H_u+\overline{H}_u
+H_d+\overline{H}_d+S\,,
\ea
\label{4}
\ee
whereas the renormalizable part of the TeV scale superpotential is given by
\be
\ba{c}
W_{B} = \lambda S (H_u H_d) + \lambda_{ij} S (H^d_{i} H^u_{j})
+ \kappa_{ij} S (D_{i} \overline{D}_{j}) + \tilde{f}_{\alpha i} S_{\alpha} (H^d_{i} H_u)
+ f_{\alpha i} S_{\alpha} (H_d H^u_{i}) \\[2mm]
+ g^{q}_{ij}\overline{D}_i d^c_4 u^c_j + h^D_{ij} d^c_4 (H^d_{i} Q_j)
+ \mu_d d^c_4\overline{d^c}_4 + \mu^u_{i} H^u_{i} \overline{H}_u
+ \mu^d_{i} H^d_{i} \overline{H}_d + W_{MSSM}(\mu=0)\,.
\ea
\label{5}
\ee
The superpotential (\ref{5}) contains a set of the TeV scale mass parameters, i.e.
$\mu_d$, $\mu^u_{i}$, $\mu^d_{i}$ that can be induced after the breakdown of local SUSY.

The gauge group and field content of the $E_6$ inspired SUSY models discussed above
can originate from the 5D and 6D orbifold GUT models in which the splitting of GUT
multiplets can be naturally achieved \cite{nevzorov}. In these orbifold GUT models
all GUT relations between the Yukawa couplings can get entirely spoiled.
On the other hand the approximate unification of the SM gauge couplings should
take place in these models. In the scenario A the analysis of the solutions of the
two--loop RG equations indicates that the gauge coupling unification can be achieved
for any phenomenologically reasonable value of $\alpha_3(M_Z)$ consistent with the
central measured low energy value \cite{nevzorov}, \cite{King:2007uj}. In the scenario B
large two--loop corrections spoil the exact unification of gauge couplings \cite{nevzorov}.
Nonetheless the relative discrepancy of $\alpha_i(M_X)$ is about 10\%  that should not
be probably considered as a big problem within the orbifold GUT framework.

\begin{table}[ht]
\centering
\begin{tabular}{|c|c|c|c|c|c|c|c|c|c|}
\hline
                   &  $27_i$          &   $27_i$              &$27'_{H_u}$&$27'_{S}$&
$\overline{27'}_{H_u}$&$\overline{27'}_{S}$&$27'_N$&$27'_{L}$&$27'_{d}$\\
& & &$(27'_{H_d})$& &$(\overline{27'}_{H_d})$& &$(\overline{27'}_N)$&$(\overline{27'}_L)$&$(\overline{27'}_{d})$\\
\hline
                   &$Q_i,u^c_i,d^c_i,$&$\overline{D}_i,D_i,$  & $H_u$     & $S$     &
$\overline{H}_u$&$\overline{S}$&$N^c_H$&$L_4$&$d^c_4$\\
                   &$L_i,e^c_i,N^c_i$ &  $H^d_{i},H^u_{i},S_i$& $(H_d)$   &         &
$(\overline{H}_d)$&&$(\overline{N}_H^c)$&$(\overline{L}_4)$&$(\overline{d^c}_4)$\\
\hline
$\tilde{Z}^{H}_2$  & $-$              & $-$                   & $+$       & $+$     &
$-$&$-$&$+$&$+$&$+$\\
\hline
$Z_{2}^{M}$        & $-$              & $+$                   & $+$       & $+$     &
$+$&$+$&$-$&$-$&$-$\\
\hline
$Z_{2}^{E}$        & $+$              & $-$                   & $+$       & $+$     &
$-$&$-$&$-$&$-$&$-$\\
\hline
\end{tabular}
\caption{Transformation properties of different components of $E_6$ multiplets
under $\tilde{Z}^H_2$, $Z_{2}^{M}$ and $Z_{2}^{E}$ discrete symmetries.}
\label{tab1}
\end{table}

The invariance of the low--energy effective Lagrangian of the $E_6$ inspired SUSY models
being considered here under both $Z_{2}^{M}$ and $\tilde{Z}^{H}_2$ symmetries implies that
it is also invariant under the transformations of $Z_{2}^{E}$ symmetry associated with exotic
states because $\tilde{Z}^{H}_2 = Z_{2}^{M}\times Z_{2}^{E}$. The transformation properties
of different components of $27_i$, $27'_l$ and $\overline{27'}_l$ supermultiplets under the
$\tilde{Z}^{H}_2$, $Z_{2}^{M}$ and $Z_{2}^{E}$ symmetries are summarized in Table~\ref{tab1}.
The $Z_{2}^{E}$ symmetry conservation implies that in collider experiments the exotic
particles, which are odd under this symmetry, can only be created in pairs and the lightest
exotic state must be stable. Using the method proposed in \cite{Hesselbach:2007te} it was
argued that the masses of the lightest and second lightest inert neutralino states ($\tilde{H}^0_1$
and $\tilde{H}^0_2$), which are predominantly the fermion components of the two SM singlet
superfields $S_i$ from $27_i$, do not exceed $60-65\,\mbox{GeV}$ \cite{Hall:2010ix}.
Since these states are odd under the $Z_{2}^{E}$ symmetry they tend to be the lightest exotic
particles in the spectrum \cite{Hall:2010ix}.

On the other hand the $Z_{2}^{M}$ symmetry conservation ensures that $R$--parity is also conserved.
Because $\tilde{H}^0_1$ is also the lightest $R$--parity odd state either the lightest $R$--parity
even exotic state or the lightest $R$--parity odd state with $Z_{2}^{E}=+1$ must be absolutely stable.
Most commonly the second stable state is the lightest ordinary neutralino $\chi_1^0$ ($Z_{2}^{E}=+1$).
Although both stable states are natural dark matter candidates in these $E_6$ inspired SUSY
models the couplings of $\tilde{H}^0_1$ to the gauge bosons, Higgs states, quarks and leptons are
rather small when $|m_{\tilde{H}^0_{1}}|\ll M_Z$. As a consequence the cold dark matter density tends
to be much larger than its measured value. In principle, $\tilde{H}^0_1$ could account for
all or some of the observed cold dark matter density if it had mass close to half the $Z$ mass
\cite{Hall:2010ix}, \cite{Hall:2009aj}. However the usual SM-like Higgs boson decays more than 95\% of
the time into either $\tilde{H}^0_1$ or $\tilde{H}^0_2$ in these cases \cite{Hall:2010ix}.
Thus the corresponding scenarios are basically ruled out nowadays.

The simplest phenomenologically viable scenarios imply that $f_{\alpha\beta}\sim \tilde{f}_{\alpha\beta}\lesssim 10^{-6}$
\cite{nevzorov}. So small values of the Yukawa couplings $f_{\alpha\beta}$ and $\tilde{f}_{\alpha\beta}$
result in extremely light inert neutralino states $\tilde{H}^0_1$ and $\tilde{H}^0_2$ which are substantially
lighter than $1\,\mbox{eV}$\footnote{The presence of very light neutral fermions in the particle spectrum
might have interesting implications for the neutrino physics (see, for example \cite{Frere:1996gb}).}.
In this case $\tilde{H}^0_1$ and $\tilde{H}^0_2$ form hot dark matter (dark radiation) in the Universe
but give only a very minor contribution to the dark matter density while the lightest ordinary neutralino
may account for all or some of the observed dark matter density.

\section{The RG flow of the gauge and Yukawa couplings}

In this section we consider the RG flow of the gauge and Yukawa couplings in the case of scenarios A and B.
The superpotential in the $E_6$ inspired SUSY models discussed in the previous section involves a lot of new
Yukawa couplings in comparison to the SM and MSSM. New couplings may be relatively large affecting the
running of all parameters. This complicates the analysis of the RG flow drastically
making it model dependent. Therefore we restrict our consideration here by the simplest
scenarios that allows to get phenomenologically viable solutions. The top--quark mass measurements
clearly indicate that the top--quark Yukawa coupling is large and can not be ignored. Nevertheless
the theoretical analysis performed in \cite{qfp-mssm1}-\cite{qfp-mssm2} revealed, that a broad class of
solutions of the MSSM RG equations concentrated near the quasi--fixed point corresponds to
$\tan\beta=1.3-1.8$. These comparatively small values of $\tan\beta$ lead to the lightest Higgs mass
which does not exceed $94\pm 5\text{~GeV}$ \cite{qfp-mssm1}. Nowadays so light SM-like Higgs
boson is ruled out. Thus in order to get phenomenologically viable solutions within the $E_6$ inspired
SUSY models studied here we allow the Yukawa coupling $\lambda$ to be as large as the top--quark
Yukawa coupling (i.e. $\lambda(M_X)\sim h_t(M_X)$). This should permit us to find self--consistent solutions
with the larger mass of the SM-like Higgs state as compared with the MSSM. Moreover large values of $\lambda$
can affect the evolution of the soft scalar mass $m_S^2$ of the singlet field $S$ rather strongly resulting in
negative values of $m_S^2$ at low energies that triggers the breakdown of $U(1)_{N}$ symmetry.
To simplify our analysis we further assumed that all other Yukawa couplings are sufficiently small so that they
can be neglected in the leading approximation. Then the approximate superpotential studied is given by:
\be
\ba{c}
W\approx \lambda S(H_{d} H_{u})+h_t (H_{u}Q_3) u^c_3\,.
\ea
\label{51}
\ee

\subsection{The running of the gauge couplings}

First of all we discuss the evolution of the SM gauge couplings $g_i(Q)$. Their values at the EW scale are
fixed by the LEP measurements and other experimental data \cite{pdg}. Assuming that the gauge coupling unification
is preserved the solutions of the one--loop RG equations for the SM gauge couplings may be
presented in the following form
\be
\frac{1}{g_i^2(Q)}=\frac{1}{g^2_0}+\frac{\beta_i}{(4\pi)^2}\ln\frac{M_X^2}{Q^2}\,,
\label{6}
\ee
where index $i$ runs from 1 to 3 and $\beta_i$ are one--loop $\beta$ functions: $\beta_1=\ds\frac{33}{5}+n_f$,
$\beta_2=1+n_f$, $\beta_3=-3+n_f$. Here $n_f$ is a number of pairs of $5+\overline{5}$ supermultiplets that
survive to the TeV scale in addition to the MSSM particle contents. In the scenarios A and B the corresponding
numbers are $n_f=3$ and $n_f=4$ respectively. Although the high energy scale $M_X$ where the unification of
the SM gauge couplings takes place is almost insensitive to $n_f$ the overall gauge coupling $g_0$ depends on
the number of exotic supermultiplets $n_f$ rather strongly. It rises when $n_f$ grows.  Indeed, in the one--loop
approximation we have
\be
\ds\frac{1}{g_0^2}=\ds\frac{1}{\beta_1-\beta_2}\biggl(\ds\frac{\beta_1}{g_2^2(M_Z^2)}-\ds\frac{\beta_2}{g_1^2(M_Z^2)}\biggr)\,.
\label{7}
\ee
For $n_f=3$ the value of the overall gauge coupling $g_0\simeq 1.2$ while for $n_f=4$ Eq.~(\ref{7}) gives $g_0\simeq 2.0$.
If $n_f > 4$ the right--hand side of Eq.~(\ref{7}) becomes negative that restricts a possible number of extra
$5+\overline{5}$ pairs which can survive to the TeV scale by four.

In the case of the $SU(2)_W$ and $U(1)_Y$ gauge couplings the large values of $g^2_0\gg 1$ imply that
the first term in the right--hand side of the Eq.~(\ref{6}) is substantially smaller than the second term and
the corresponding solutions of the RG equations (RGEs) are focused near the infrared stable fixed point
at low energies, i.e.
\be
\ds\frac{g_1^2}{g_2^2}\simeq\frac{\beta_2}{\beta_1}\,.
\label{8}
\ee
This fixed point corresponds to $\ds\frac{d g_1/g_2}{dt}\to 0$, where $t=\ln\left(M_X/Q\right)$, $Q$ is a renormalisation
scale. In general the solutions of the RG equations always approach the infrared stable fixed point when $t\to \infty$.
In our analysis the interval of variations of $t$ remains always finite, i.e. $0\le t\le \ds\ln\frac{M_X^2}{M_Z^2}$.
As a consequence the solutions for $g_i(Q)$ are concentrated near the quasi--fixed points  which set upper limits on the
allowed range of the low-energy values of these couplings caused by the applicability of perturbation theory up to the scale
$M_X$, i.e. $g_0\lesssim \sqrt{4\pi}$.

In the case of scenarios A and B the values of the $SU(2)_W$ and $U(1)_Y$ gauge couplings calculated in the limit $g^2_0\gg 1$
are relatively close to the measured values of these couplings, i.e. $g_1 (M_Z)\simeq 0.461$ and $g_2 (M_Z)\simeq 0.652$.
The ratio of the measured values of the $SU(2)_W$ and $U(1)_Y$ gauge couplings $g_1^2 (M_Z)/g_2^2 (M_Z)\simeq 0.5$
whereas Eq.~(\ref{8}) gives $\ds\frac{g_1^2}{g_2^2}\simeq 0.47$ in the scenario B and $\ds\frac{g_1^2}{g_2^2}\simeq 0.42$
in the scenario A\footnote{In the MSSM the infrared fixed point (\ref{8}) is very far from the corresponding ratio of the physical
quantities of the $SU(2)_W$ and $U(1)_Y$ gauge couplings. Indeed, for $n_f=0$ Eq.~(\ref{8}) gives $\ds\frac{g_1^2}{g_2^2}\simeq 0.15$}.

If $\beta_3 > 0$, like in the scenario B, the solutions of the RG equations for the $SU(2)_W$ and $SU(3)_C$ gauge couplings
also approach the fixed point
\be
\ds\frac{g_2^2}{g_3^2}\simeq\frac{\beta_3}{\beta_2}
\label{9}
\ee
in the limit $g^2_0\to \infty$. However since even in the scenario B the value of the one--loop beta function associated with the strong
interactions is rather small, i.e. $\beta_3=1$,  the convergence of the solutions for $g_3(Q)$ to the corresponding quasi--fixed point
is rather weak. Therefore the solutions of the RG equations for $g_2(Q)/g_3(Q)$ are also attracted to the fixed point  (\ref{9})
very weakly. In the case of scenario A $\beta_3$ vanishes in the one--loop approximation so that near the fixed point  (\ref{9})
$\ds\frac{g_2^2}{g_3^2}\to 0$. It means that the ratios $g_{1,2}^2/g_3^2$ become extremely small when $t\to \infty$. At any
low-energy scale $Q$ the value of the strong gauge coupling in the scenarios A and B is substantially larger than $g_1(Q)$ and
$g_2(Q)$ so that the ratios $g_{1,2}^2/g_3^2$ are quite small but not negligible.

The inclusion of the two--loop corrections shifts the position of the quasi--fixed points where the solutions of the RG equations
are focused. The values of the gauge couplings at the EW scale calculated for different values of $g_0$ in the two--loop
approximation are given in Table~\ref{tab2}. The two--loop RG flow of gauge couplings is shown in Fig.~1. The corresponding two--loop
beta functions can be found in \cite{nevzorov}. The results presented in Table 2 demonstrate that the inclusion of the
two--loop corrections leads to the growth of the ratio $\ds\frac{g_1^2}{g_2^2}$ near the quasi--fixed points. Indeed,
for $g_0=3$ in the scenario A $\ds\frac{g_1^2}{g_2^2}$ increases from $0.43$ to $0.48$ whereas in the scenario B
$\ds\frac{g_1^2}{g_2^2}$ grows from $0.48$ to $0.52$.

In the case of scenario A the typical pattern of the RG flow of the gauge couplings from $Q=M_X$ to the EW scale for
different values of $g_0$ is presented in Fig.~1. The same plots can be obtained in the scenario B as well. Since plots in the
case of scenarios A and B look very similar we include only ones that correspond to the scenario A. From Fig.~1a it follows that
the solutions of the RG equations for $g_1(Q)$ and $g_2(Q)$ are sufficiently strongly attracted to the quasi--fixed points.
In the scenarios A and B the numerical values of these gauge couplings associated with the quasi-fixed points (see Table~\ref{tab2})
are reasonably close to the measured values of these couplings. On the other hand as one can see from Fig.~1b
the convergence of the solutions of the RG equations for $g_3(Q)$ to the quasi--fixed point is rather weak.  Moreover
our numerical analysis reveals that in the case of scenario B the values of $g_3(Q)$ at the EW scale tend to be substantially
smaller than the experimentally measured central value of this coupling\footnote{In the scenario B the considerably larger values
of the strong gauge coupling at the EW scale can be obtained if we take into account the low energy threshold effects associated
with the presence of exotic states and superpartners of ordinary particles. Nevertheless even in this case the exact unification of
gauge couplings can be achieved only for $\alpha_3(M_Z)\lesssim 0.112$. For $\alpha_3(M_Z)\simeq 0.118$ the relative
discrepancy of $\alpha_i(M_X)$ is about 10\%.}. In the scenario A the value of $g_3(Q)$, where the solutions of the RG equations
are focused at low energies, is considerably larger than the one that corresponds to $\alpha_3(M_Z)\simeq 0.118$. At the same time
the results presented in Table~\ref{tab2} indicate that for $g_0=1.5$ all SM gauge couplings at the EW scale including the strong
gauge coupling are rather close to their measured central values in the case of scenario A.

The RG flow of the gauge couplings in the scenarios A and B is affected by the kinetic term mixing which is consistent with all
symmetries. Indeed, in the most general case the gauge kinetic part of the Lagrangian can be written as
\be
\mathcal{L}_{kin}=-\ds\frac{1}{4}\left(F^Y_{\mu\nu}\right)^2-
\frac{1}{4}\left(F^{N}_{\mu\nu}\right)^2-\frac{\sin\chi}{2}F^{Y}_{\mu\nu}F^{N}_{\mu\nu}\,-...,
\label{10}
\ee
where $F_{\mu\nu}^Y$ and $F^{N}_{\mu\nu}$ are field strengths for the $U(1)_Y$ and $U(1)_N$ gauge interactions,
while $B^{Y}_{\mu}$ and $B^{N}_{\mu}$ are the corresponding gauge fields respectively. In Eq.~(\ref{10}) the
terms associated with the $SU(3)_C$ and $SU(2)_W$ gauge interactions are omitted. If $U(1)_Y$ and $U(1)_N$ symmetries
arise from the breaking of the simple gauge group $E_6$ the parameter $\sin\chi$ which parametrizes the gauge kinetic term
mixing is expected to vanish near the GUT scale. Nevertheless it gets induced due to loop effects since
\be
Tr\left(Q^YQ^{N}\right)=\sum_{i=chiral fields}\left(Q^Y_i Q^{N}_i\right)\ne 0\,.
\label{11}
\ee
Here the trace is restricted to the states lighter than the energy scale being considered. The complete $E_6$ multiplets do
not contribute to the trace (\ref{11}). Its non--zero value is caused by the presence of the components of the incomplete
$27'_l$ and $\overline{27'}_l$ multiplets of the original $E_6$ symmetry which survive to low energy.

The mixing in the gauge kinetic part of the Lagrangian (\ref{10}) can be eliminated by a non--unitary transformation of
two $U(1)$ gauge fields \cite{33}:
\be
B^Y_{\mu}=B_{1\mu}-B_{2\mu}\tan\chi\,,\qquad B^{N}_{\mu}=B_{2\mu}/\cos\chi\,.
\label{12}
\ee
In the new basis of the gauge fields $(B_{1\mu},\,B_{2\mu})$  the gauge kinetic part of the Lagrangian is diagonal
whereas the covariant derivative can be written in a compact form
\be
D_{\mu}=\partial_{\mu}-iQ^{T}GB_{\mu}\,...,
\label{13}
\ee
where $Q^T=(Q^Y_i,\,Q^{N}_i)$, $B^{T}_{\mu}=(B_{1\mu},\,B_{2\mu})$ and $G$ is
a $2\times 2$ matrix of gauge couplings
\be
G=\left(
\ba{cc}
g_1 & g_{11}\\[2mm]
0   & g'_1
\ea
\right)\,, \qquad g_1=g_Y\,, \qquad g'_1=g_{N}/\cos\chi\,,\qquad g_{11}=-g_Y\tan\chi\,.
\label{14}
\ee
In the expression for the covariant derivative (\ref{13}) the $SU(3)_C$ and $SU(2)_W$ gauge fields
are omitted. In Eq.~(\ref{14}) $g_Y$ and $g_{N}$ are original $U(1)_Y$ and $U(1)_N$ gauge couplings
which are supposed to be equal at the scale $M_X$. In the considered approximation the gauge kinetic
mixing changes effectively the $U(1)_N$ charges of the fields to $\tilde{Q}^{N}_i=Q^{N}_i+Q^{Y}_i\delta$, where
$\delta=g_{11}/g'_1$ while the $U(1)_Y$ charges remain the same.

Using the matrix notation for the structure of $U(1)$ interactions one can write down the RG
equations for the Abelian couplings in a compact form  \cite{34}:
\be
\ds\frac{d G}{d t}=-G\times B\,, \qquad\qquad
B=\ds\frac{1}{(4\pi)^2}
\left(
\ba{cc}
\beta_1 g_1^2 & 2g_1g'_1\beta_{11}+2g_1g_{11}\beta_1\\[2mm]
0 & g^{'2}_1\beta'_1+2g'_1 g_{11}\beta_{11}+g_{11}^2\beta_1
\ea
\right)\,.
\label{15}
\ee
From Eqs.~(\ref{15}) one can see that, whereas the solution of the one--loop RG equation
for $g_1(Q)$ is still described by Eq.~(\ref{6}), the running of couplings $g'_1(Q)$ and $g_{11}(Q)$
obey quite complicated system of differential equations. In the scenario A $\beta'_1=47/5$ and
$\beta_{11}=-\sqrt{6}/5$ in the one--loop approximation. In the case of scenario B the one--loop
$\beta'_1$ and $\beta_{11}$ are $10.9$ and $\beta_{11}=3\sqrt{6}/10$ respectively.

In the $E_6$ inspired SUSY models with extra $U(1)_N$ factor the RG equations (\ref{15})
have infrared stable fixed points:
\be
\ds\frac{g_{11}}{g'_1}=-\frac{\beta_{11}}{\beta_1}\,,\qquad
\ds\frac{g_1^2}{g^{'2}_1}=\frac{\beta'_1}{\beta_1}-\left(\frac{\beta_{11}}{\beta_1}\right)^2\,.
\label{16}
\ee
The solutions of the differential equations (\ref{15}) approach the fixed points (\ref{16}) when the overall
gauge coupling $g_0$ and $t$ increase. Since in both scenarios $\beta_1 \simeq \beta'_1\gg \beta_{11}$
the values of the diagonal $U(1)_Y$ and $U(1)_N$ gauge couplings are approximately equal at low energies
whereas the off--diagonal gauge coupling $g_{11}(Q)$ being set to zero at the GUT scale remains rather
small at any scale below $M_X$. Eq~(\ref{16}) indicates~ that~ in~ the~ case~ of~ scenario~ A~ $g_1$~
tends~ to~ be~ slightly~ less~ than~ $g'_1$~ near~ the~ fixed~ point~ while~ in~ the~ scenario~ B~
$g_1\gtrsim g'_1$.

The two--loop RG flow of $g^{'2}_1/g_2^2$ and $g_{11}/g_2$ are shown in Fig.~1c and 1d where
we set $g'_1(M_X)=g_0$ and $g_{11}(M_X)=0$. Because $g_{11}(Q) \ll g_i(Q)$ and $\beta_{11}$
is relatively small as compared with the diagonal beta functions we neglect two--loop corrections
to $\beta_{11}$. Again we only include plots associated with the scenario A because the corresponding
plots look rather similar in both scenarios. One can see that Figs.~1a and 1c are almost identical.
This is because $g_1(Q)\simeq g'_1(Q)$. The results presented in Fig.~1 and Table~\ref{tab2} demonstrate
that the inclusion of the two--loop corrections don't change much the position of the fixed
points (\ref{16}). In principle the two--loop corrections to $\beta_3$, $\beta_2$, $\beta_1$ and $\beta'_1$
as well as the two--loop RG flow of all gauge couplings depend on $h_t(Q)$ and $\lambda(Q)$.
However this dependence is rather weak and can be ignored in the first approximation \cite{King:2007uj}.
Nevertheless the results presented in Table~\ref{tab2} and Fig.~1 are obtained for $h_t(M_X)=\lambda(M_X)=g_0$.
The evolution of the Yukawa couplings will be considered in the next subsection.

\subsection{The running of the Yukawa couplings and the Higgs mass}

Since the RG flow of gauge couplings in the $E_6$ inspired SUSY models with extra $U(1)_N$ factor
implies that the corresponding quasi--fixed points of RG equations are reasonably close to the
measured values of $g_i(M_Z)$  it is worth to examine the quasi--fixed point solutions for the
Yukawa couplings as well. The Yukawa couplings appearing in the superpotential (\ref{51}) obey
the following two--loop RG equations:
\be
\ba{rcl}
\ds\frac{d\lambda}{dt}&=&\ds\frac{\lambda}{(4\pi)^2}\biggl[-4\lambda^2 -3h_t^2 +3g_2^2+ \dfrac{3}{5}g_1^2 +\dfrac{19}{10} g^{'2}_1
-\dfrac{1}{(4\pi)^2}\biggl\{-10\lambda^4 - 9\lambda^2 h_t^2 - 9 h_t^4 \\[0mm]
&+& \lambda^2 \left(6 g_2^2 + \dfrac{6}{5} g_1^2 + \dfrac{13}{10} g_1^{'2}\right)
+ h_t^2 \left(16 g_3^2 + \dfrac{4}{5} g_1^2 - \dfrac{3}{10} g_1^{'2}\right)+ b_{\lambda} g_2^4 + c_{\lambda} g_1^4 \\[0mm]
&+& d_{\lambda} g_1^{'4} + \ds\frac{9}{5} g_2^2 g_1^2 + \ds\frac{39}{20} g_2^2 g_1^{'2} + \ds\frac{39}{100} g_1^2 g_1^{'2}\biggr\}\biggr]\,,\\[2mm]
\ds\frac{dh_t}{dt}&=&\ds\frac{h_t}{(4\pi)^2}\biggl[-\lambda^2-6h_t^2+\ds\frac{16}{3}g_3^2+3g_2^2+\ds\frac{13}{15}g_1^2+
\frac{3}{10} g^{'2}_1 - \dfrac{1}{(4\pi)^2}\biggl\{-3\lambda^4  \\[0mm]
& - &3\lambda^2 h_t^2 -22 h_t^4 + \dfrac{3}{2}\lambda^2 g_1^{'2} +  h_t^2 \left(16 g_3^2 + 6 g_2^2 + \dfrac{6}{5} g_1^2 +
\dfrac{3}{10} g_1^{'2}\right) + a_{h_t} g_3^4 \\[0mm]
& + &b_{h_t} g_2^4 + c_{h_t} g_1^4 +d_{h_t} g_1^{'4} + 8g_3^2g_2^2+\ds\frac{136}{45}g_3^2g_1^2
+\ds\frac{8}{15}g_3^2g_1^{'2}+g_2^2g_1^2\\[0mm]
& + & \ds\frac{3}{4}g_2^2g_1^{'2}+\ds\frac{53}{300}g_1^2g_1^{'2}
\biggr]\,,
\ea
\label{17}
\ee
where in the case of scenario A
\be
\ba{llll}
a_{\lambda}=0\,,\qquad & b_{\lambda}=\dfrac{33}{2}\,,\qquad & c_{\lambda}=\dfrac{297}{50}\,,\qquad &d_{\lambda}=\dfrac{3933}{200}\,,\\[3mm]
a_{h_t}=\dfrac{128}{9}\,,\qquad& b_{h_t}=\dfrac{33}{2}\,,\qquad & c_{h_t}=\dfrac{3913}{450}\,,\qquad &d_{h_t}=\dfrac{573}{200}\,,
\ea
\label{18}
\ee
while in the scenario B we have
\be
\ba{llll}
a_{\lambda}=0\,,\qquad & b_{\lambda}=\dfrac{39}{2}\,,\qquad & c_{\lambda}=\dfrac{327}{50}\,,\qquad &d_{\lambda}=\dfrac{4503}{200}\,,\\[3mm]
a_{h_t}=\dfrac{176}{9}\,,\qquad& b_{h_t}=\dfrac{39}{2}\,,\qquad & c_{h_t}=\dfrac{4303}{450}\,,\qquad &d_{h_t}=\dfrac{663}{200}\,.
\ea
\label{19}
\ee
In the right--hand side of Eq.~(\ref{17}) we neglect all Yukawa couplings except $\lambda$ and $h_t$.

From Eq.~(\ref{17}) it follows that the evolution of $\lambda(Q)$ and $h_t(Q)$ depend on the values of the gauge couplings.
In the case of scenario A we set $g_0=1.5$. As it was pointed out in the previous subsection this value of the overall gauge
coupling leads to $g_i(M_Z)$ which are very close to their measured central values. In the scenario B we fix $g_0=3$ because
it results in the most phenomenologically acceptable values of gauge couplings at low energies.

For the purposes of our RG studies, it is convenient to introduce
\be
\rho_t=\ds\frac{h_t^2}{g_3^2}\,,\qquad\qquad  \rho_{\lambda}=\ds\frac{\lambda^2}{g_3^2}\,.
\label{20}
\ee
The allowed range of the parameter space in the $(\rho_t, \rho_{\lambda})$ plane is limited at the EW scale by the
quasi--fixed (or Hill type effective) line. Outside this range the solutions for $h_t(Q)$ and $\lambda(Q)$
develop a Landau pole below the scale $M_X$ so that the perturbation theory becomes inapplicable. The
solutions of the RG equations (\ref{17}) are gathered near this line, when the Yukawa couplings at the GUT
scale $M_X$ increase. However the allocation of the solutions for $\rho_t$ and $\rho_{\lambda}$ at the EW
scale along the Hill line is not uniform. The main reason for this is that at large values of the Yukawa couplings
at the scale $M_X$ the corresponding solutions are attracted not only to the Hill line but also to the invariant
(or infrared fixed) line. When $t$ goes to zero, this line approaches its asymptotic limit where
$\rho_t, \rho_{\lambda}>>1$ and $\rho_{\lambda}\to \rho_t$, which is a fixed point of the RG equations
for the Yukawa couplings in the gaugeless limit ($g_1=g_2=g_3=g'_1=0$). The invariant line connects this
fixed point with the infrared stable fixed point. In the scenario A this fixed point is given by:
\be
\rho_{\lambda}=0\,, \qquad \rho_t\simeq 0.89.
\label{21}
\ee
whereas in the scenario B
\be
\rho_{\lambda}=0\,, \qquad \rho_t\simeq 1.17.
\label{22}
\ee
All solutions of the RG equations for $\rho_t$ and $\rho_{\lambda}$ are concentrated near the infrared stable fixed point
at very low energies when $t\to\infty$. The infrared fixed line is RG invariant solution. If the boundary values at $Q=\Lambda$
are such that $h_t(\Lambda)$ and $\lambda(\Lambda)$ belong to the fixed line, the evolution of the Yukawa couplings proceeds
further along this line towards the infrared stable fixed point. With increasing of the interval of the RG flow the solutions
of the differential equations (\ref{17}) are first attracted to the invariant line and then close to or along this line towards
the infrared fixed point. Infrared fixed lines and surfaces, as well as their properties, were studied in detail in \cite{schrempp}.

As $h_{t}(M_X)$ and $\lambda(M_X)$ grow, the region at the EW scale in which the solutions of the RG equations
for $\rho_t$ and $\rho_{\lambda}$ are concentrated shrinks drastically. They are focused near the intersection point of
the invariant and quasi--fixed lines. Hence this point can be considered as the quasi--fixed point of the
RG equations (\ref{17}) \cite{nmssm-qfp}. In the two--loop approximation the intersection points of
the invariant and quasi--fixed lines have the following coordinates in the $(\rho_t, \rho_{\lambda})$ plane
\be
(A)~\rho_t=1.16\,,\quad\rho_{\lambda}=0.14;\qquad\qquad (B)~\rho_t=1.33\,,\quad\rho_{\lambda}=0.18\,.
\label{23}
\ee
in the cases of scenarios A and B respectively. The quasi--fixed points (\ref{23}) correspond to
$h_t(M_X)=\lambda(M_X)\simeq 3$, i.e. $\rho_{\lambda}(M_X)=\rho_t(M_X)$, which is associated with the
fixed point of the RG equations for the Yukawa couplings in the gaugeless limit. Eq.~(\ref{23}) indicates
that turning the gauge couplings on induces a certain hierarchy between $h_t(Q)$ and $\lambda(Q)$.
Indeed, because $g_3(Q)$ is substantially larger than other gauge couplings at low energies the
top--quark Yukawa coupling tends to dominate over $\lambda(Q)$.

The two--loop RG flow of $\rho_t(Q)$ and $\rho_{\lambda}(Q)$ in the cases of scenarios A and B are
shown in Figs.~2a and 2b. The results of our analysis are also summarised in Table~\ref{tab3}. In Figs.~2a and
2b we plot the running $\rho_{\lambda}(Q)$ versus $\rho_t(Q)$ from $Q=M_X$ to the EW scale
for regular distribution of boundary conditions for $\lambda(M_X)$ and $h_t(M_X)$ at the GUT scale.
These plots demonstrate that the trajectories, which represent different solutions of the two--loop RG
equations are focused in a narrow region near the quasi--fixed points at low energies. From Table~\ref{tab3}
it follows that the relative variations of $h_t(M_Z)$ near the quasi fixed point are rather small,
i.e. about 1\%, when $1.5\lesssim h_t(M_X),\,\lambda(M_X)\lesssim 3$. The interval of variations of
$\lambda(M_Z)$ is substantially wider. The relative deviations of $\lambda(M_Z)$ can be as large as
20 percents when $h_t(M_X)$ and $\lambda(M_X)$ vary from 1.5 to 3. As one can see from
Fig.~2a in the scenario A different trajectories also tend to flow towards the invariant line that
corresponds to $\rho_{\lambda}(M_X)=\rho_t(M_X)=4$. This is less obvious in the case of scenario B
since $g_0=3$ and the Yukawa couplings $h_{t}(M_X)$ and $\lambda(M_X)$ have to be either of
order of or even smaller than the gauge ones to ensure the validity of perturbation theory. Thus the
gaugeless approximation is inapplicable.

The convergence of $h_t(Q)$ to the quasi--fixed points (\ref{23}) allows to predict the value of the
top quark Yukawa coupling at the EW scale. Then, using the relation between the running mass and
Yukawa coupling of the $t$--quark
\begin{equation}
m_t(M_t)=\dfrac{h_t(M_t)}{\sqrt{2}}v\sin\beta,
\label{24}
\end{equation}
one can find the value of $\tan\beta$ that corresponds to the quasi--fixed points (\ref{23}).
In Eq.~(24) $v=\sqrt{v_1^2+v_2^2}=246\,\mbox{GeV}$, $\tan\beta=v_2/v_1$ while
$v_2$ and $v_1$ are the VEVs that Higgs doublets $H_u$ and $H_d$ develop. In our analysis
we set $m_t(M_t)\simeq 163\,\mbox{GeV}$ which is rather close to the central value
that can be obtained using the world average mass of the top quark $M_t=173.07\pm 0.52\pm 0.72$ GeV
(see \cite{pdg}) and the relationship between the $t$--quark pole ($M_t$) and running ($m_t(Q)$) masses \cite{mtMS}
\begin{equation}
m_t(M_t)=M_t\biggl[1-1.333\ds\,\frac{\alpha_s(M_t)}{\pi}-
9.125\left(\ds\frac{\alpha_s(M_t)}{\pi}\right)^2\biggr]\,.
\label{25}
\end{equation}
From Table~\ref{tab3} one can see that $\tan\beta=1.02-1.05$ in the scenario A and $\tan\beta=1.19-1.22$
in the scenario B when $h_t(M_X)$ and $\lambda(M_X)$ vary from 1.5 to 3.

The spectrum of the Higgs bosons in the $E_6$ inspired SUSY models with extra $U(1)_N$ factor involves a set of
the neutral Higgs states. Like in the MSSM, one of the neutral CP--even Higgs states, which manifests
itself in the interactions with gauge bosons and fermions as a SM--like Higgs boson, is always light irrespective of
the SUSY breaking scale. In the leading approximation two--loop upper bound on the mass of the lightest Higgs
particle in the $E_6$ inspired SUSY models with extra $U(1)_N$ symmetry can be written as \cite{King:2005jy}
\be
\ba{c}
m_{h_1}^2\le\biggl[\ds\frac{\lambda^2}{2}v^2\sin^22\beta+M_Z^2\cos^22\beta+g^{'2}_1v^2\biggl(\tilde{Q}_1\cos^2\beta+
\tilde{Q}_2\sin^2\beta\biggr)^2\biggr]\times\qquad\qquad\\[5mm]
\times\left(1-\ds\frac{3h_t^2}{8\pi^2}l\right)
+\ds\frac{3 h_t^4 v^2 \sin^4\beta}{8\pi^2}\left\{\ds\frac{1}{2}
U_t+l+\ds\frac{1}{16\pi^2}\biggl(\frac{3}{2}h_t^2-8g_3^2\biggr)(U_t+l)l\right\}\,,
\ea
\label{26}
\ee
$$
U_t=2\ds\frac{X_t^2}{M_S^2}\biggl(1-\frac{1}{12}\frac{X_t^2}{M_S^2}\biggr)\,,\qquad\qquad\qquad l=\ln\biggl[\ds\frac{M_S^2}{m_t^2}\biggr]\,,
$$
where $\tilde{Q}_1$ and $\tilde{Q}_2$ are effective $U(1)_N$ charges of $H_d$ and $H_u$ respectively,
$X_t$ is a stop mixing parameter, $M_S$ is a SUSY breaking scale defined as $m_Q^2=m_U^2=M_S^2$ and $m_Q^2$, $m_U^2$ are soft
scalar masses of superpartners of the left--handed and right--handed components of the $t$--quark. Eq.(\ref{26}) is a simple generalization of the
approximate expressions for the upper bounds on the lightest Higgs boson mass obtained in the MSSM \cite{mh-mssm} and next-to-minimal
supersymmetric standard model (NMSSM) \cite{mh-nmssm}. At $\lambda=0$ and $g'_1=0$ the right--hand side of Eq.~(\ref{26}) coincide
with the theoretical bound on the lightest Higgs mass in the MSSM which does not exceed Z-boson mass ($M_Z\simeq 91.2\,\mbox{GeV}$)
at the tree--level \cite{23}. Leading one--loop and two--loop corrections to $m_{h_1}$ increase the upper bound on the lightest Higgs boson
mass from $M_Z$ to $130\,\mbox{GeV}$ (see \cite{Djouadi:2005gj} and references therein).  In the MSSM the approximate
expression (\ref{26}) leads to the value of the lightest Higgs mass which is typically a few $\mbox{GeV}$ lower than the one which is computed
using the Suspect \cite{suspect} and FeynHiggs \cite{feynhiggs} packages.

In our analysis we focus on the so-called maximal mixing scenario, when $X_t=\sqrt{6} M_S$, that leads to the maximal possible value of $m_{h_1}$.
We also set stop scalar masses to be equal to $m_Q=m_U=M_S=1200\,\mbox{GeV}$ that result in the reasonably light stops which are not ruled by
the LHC experiments. Then for each set of $\lambda(M_Z)$ and $\tan\beta$ one can calculate the theoretical restriction on $m_{h_1}$.
The analysis performed in \cite{King:2005jy} shows that in this case the two--loop upper bound on the lightest Higgs mass reaches its maximal
value, i.e. $150-155\,\mbox{GeV}$, for $\tan\beta\simeq 1.5-2$ when the low energy value of the coupling $\lambda$ can be as large as $0.7-0.8$\,.
The results presented in Table~\ref{tab3} indicate that the quasi--fixed point solutions in the scenarios A and B correspond to substantially lower values of $\tan\beta$
and $\lambda(M_Z)$. Therefore the two--loop upper bound on the lightest Higgs mass is also considerably lower than $155\,\mbox{GeV}$. On the other hand
in order to get solutions which can be consistent with the observation of the SM--like Higgs state with mass around $\sim 125\mbox{GeV}$
the values of the coupling $\lambda(M_Z)$ should be larger than $g'_1(M_Z)$ at least. From Table~\ref{tab3} one can see that it is possible to
find such solutions in the vicinity of the quasi--fixed point in the case of the scenario A. In the scenario B  the low energy values of $\lambda(M_Z)$
are typically smaller than the ones in the scenario A and  it seems to be rather problematic to find phenomenologically acceptable solutions near
the corresponding quasi--fixed point.

The requirement that $\lambda\gtrsim g'_1$ at the EW scale leads to extremely hierarchical structure of the Higgs spectrum \cite{King:2005jy}.
Indeed, in this case the qualitative pattern of the Higgs spectrum is rather similar to the one that arises in the PQ symmetric NMSSM
in which the heaviest CP--even, CP--odd and charged states are almost degenerate and much heavier than the lightest and second
lightest CP-even Higgs bosons \cite{Miller:2005qua}--\cite{Nevzorov:2004ge}. Because the mass of the second lightest CP--even
Higgs state is set by the $Z'$ boson mass ($M_{Z'}$) \cite{King:2005jy} which should be heavier than $2\,\mbox{TeV}$ the heaviest
Higgs boson masses lie beyond the multi TeV range and the mass matrix of the CP--even Higgs sector can be diagonalized using
the perturbation theory \cite{Nevzorov:2004ge}--\cite{Nevzorov:2001um}. Thus the phenomenologically viable quasi--fixed point solutions
in the $E_6$ inspired SUSY models with extra $U(1)_N$ gauge symmetry implies that all Higgs states except the lightest one are
extremely heavy and can not be discovered at the LHC. In this limit the lightest CP--even Higgs boson is the analogue of the SM Higgs
field. Extremely hierarchical structure of the Higgs spectrum also implies that all phenomenologically viable quasi--fixed point scenarios
are quite fine--tuned.

It is useful to compare the results of our analysis of the quasi--fixed point scenarios in the $E_6$ inspired SUSY models with the
corresponding results in more simple SUSY extensions of the SM like the NMSSM \cite{NMSSM1} and its modifications
\cite{NMSSM3}. In the NMSSM, the spectrum of the MSSM is extended by one singlet superfield (for reviews
see~\cite{NMSSM2}). The term $\mu (H_d H_u)$ in the superpotential is then replaced by the coupling term
$\lambda S H_d H_u$. As in the $E_6$ inspired SUSY models discussed above the superfield $S$ acquires non-zero VEV
($\langle S \rangle =s/\sqrt{2}$) and an effective $\mu$-term ($\mu_{eff}=\lambda s/\sqrt{2}$) is automatically generated.
However the simplest model of this type possesses an extended global $SU(2)\times [U(1)]^2$ symmetry\footnote{In the MSSM
this global symmetry of the Lagrangian reduces to the gauge one because of the $\mu$-term in the superpotential.} that after its
breakdown leads to the appearance of the massless CP-odd scalar particle in the Higgs boson spectrum which is a
Peccei--Quinn axion \cite{12}. The common way to avoid axion is to introduce a term cubic in the new singlet superfield
$\dfrac{\kappa}{3} S^3$ in the superpotential that explicitly breaks an additional $U(1)$ global symmetry. Here to simplify
our analysis of the RG flow of the Yukawa couplings we assume that $\kappa$ is negligibly small while the extended global
$SU(2)\times [U(1)]^2$ symmetry is explicitly broken by some other mechanism like in some modifications of the NMSSM
\cite{NMSSM3}.

The approximate analytical expression (\ref{26}) can be used for the calculation of the upper bound on the lightest
Higgs mass $m_{h_1}$ in the NMSSM and its modifications if we set $g'_1=0$. From Eq.~(\ref{26}) it follows that
for large $\lambda$, i.e. $\lambda>\sqrt{2} M_Z/v\simeq 0.52$, the theoretical restriction on $m_{h_1}$
attains its maximal value for $\tan\beta\sim 1$ which is larger than the upper bound on the mass of the lightest Higgs
boson in the MSSM. As a consequence for large low energy values of $\lambda$ the fine-tuning of the MSSM, which
is needed to ensure that this model is consistent with $125-126\,\mbox{GeV}$ SM--like Higgs boson, can be ameliorated
within the NMSSM \cite{NMSSMtuning}. However in the NMSSM $\lambda(M_Z) \simeq 0.7$ is the largest value in order not
to spoil the validity of perturbation theory up to the scale $M_X$. The inclusion of extra $5+\overline{5}$--plets of matter
enlarges the allowed range of $\lambda$ at low energies \cite{Masip:1998jc}. In this context we also explore here the RG flow
of the Yukawa couplings within the NMSSM with three families of $5+\overline 5$--plets of  extra matter (NMSSM+)
\cite{nmssm1+}-\cite{nmssm2+} assuming again that the coupling $\kappa$ is so small that it can be ignored
in the leading approximation.

The results of our numerical analysis of the two--loop RG flow of the Yukawa couplings are presented in Fig.~2c and 2d
as well as in Table~\ref{tab4}. The complete set of the two--loop RG equations that describe the running of the gauge and
Yukawa couplings from $Q=M_X$ to the EW scale within the NMSSM and NMSSM+ can be found in \cite{nmssm1+}.
In the case of the NMSSM we set $g_0=0.725$ whereas for the analysis of the RG flow of $h_t(Q)$ and $\lambda(Q)$
within the NMSSM+ we fix $g_0=1.5$. These values of $g_0$ leads to $g_i(M_Z)$ which are very close to the experimentally
measured values of the SM gauge couplings at the EW scale. Fig.~2c and 2d demonstrate that different trajectories
associated with different solutions of the two--loop RG equations tend to get attracted to the invariant line, that
corresponds to $\rho_{\lambda}\simeq \rho_{t}$ at high energies, and focused in a relatively narrow region at low
energies. In the $(\rho_t, \rho_{\lambda})$ plane the intersection points of the invariant and quasi--fixed line have
the following coordinates
\be
(C)~\rho_t=0.80\,,\quad\rho_{\lambda}=0.19;\qquad\qquad (D)~\rho_t=1.15\,,\quad\rho_{\lambda}=0.14
\label{27}
\ee
in the cases of the NMSSM and NMSSM+ respectively.

Naively, one can expect that the inclusion of extra $5+\overline{5}$--plets of  matter should lead to the
larger values of the Yukawa couplings at low energies. Indeed, as it was mentioned before extra multiplets of matter
change the running of the SM gauge couplings so that their values at the intermediate scale rise when the number
of new supermultiplets increases. Since $g_i(Q)$ occurs in the right--hand side of the RG equations with negative
sign the growth of the gauge couplings prevents the appearance of the Landau pole in the evolution of the Yukawa
couplings. It means that in the NMSSM+ $\lambda(M_Z)$ and $h_t(M_Z)$ are allowed to be larger than in
the NMSSM so that the lightest Higgs boson in the NMSSM+ can be heavier than in the NMSSM and MSSM.
On the other hand the results presented in Table~\ref{tab4} and the coordinates of the quasi--fixed points (\ref{27})
indicate that the values of $\lambda(M_Z)$ near the quasi--fixed point (D) tend to be smaller than in the
vicinity of quasi--fixed point (C). Thus for a fixed set of the Yukawa couplings at the GUT scale the theoretical
restrictions on the mass of the SM--like Higgs boson becomes even more stringent after the inclusion of exotic
supermultiplets of matter (see Table~\ref{tab4}). This happens because $h_t(Q)$ renormalizes by means of strong interactions
while $\lambda(Q)$ does not. Due to this the top--quark Yukawa coupling rises significantly (see Table~\ref{tab4}) resulting in
decreasing of $\lambda(Q)$ which prevails the growth of this coupling caused by the larger values of $g_i(Q)$.
While the increase of the top--quark Yukawa coupling at the EW scale leads to the decreasing of $\tan\beta$, that
pushes the lightest Higgs boson mass up, the decreasing of $\lambda(M_Z)$ reduces the upper limit on $m_{h_1}$.
The results of the numerical analysis collected in Table~\ref{tab4} show that for a fixed set of $h_t(M_X)$ and $\lambda(M_X)$
last effect dominates. As a consequence the upper bound on $m_{h_1}$ in the vicinity of the quasi--fixed point (D)
tend to be substantially smaller than $125\,\mbox{GeV}$ so that the corresponding quasi--fixed point scenario
in the NMSSM+ is basically ruled out. In the NMSSM near the quasi--fixed point (C) the low energy values of the
top--quark Yukawa coupling are smaller while $\lambda(M_Z)$ and $\tan\beta$ are larger than the ones that
correspond to the quasi--fixed point scenarios (A), (B) and (D). As a result it seems to be possible to find  in the vicinity
of the quasi--fixed point (C) phenomenologically acceptable solutions with $125-126\,\mbox{GeV}$ SM--like Higgs boson.

As in the quasi--fixed point scenarios (A) and (B)  the relative variations of $h_t(M_Z)$ near the quasi--fixed points (C)
and (D) are quite small, i.e. about 4\% and 1\% respectively when $h_t(M_X)$ and $\lambda(M_X)$ vary from 1.5 to 3
(see Table~\ref{tab4}). As before the relative deviations of $\lambda(M_Z)$ can be substantially larger, i.e. about 10\%-20\%
for the same interval of variations of $h_t(M_X)$ and $\lambda(M_X)$. Moreover the values of the gauge and Yukawa
couplings as well as $\tan\beta$ associated with the quasi--fixed points (A) and (D) are rather close. At the same time the upper
bound on $m_{h_1}$ in the the $E_6$ inspired SUSY model with extra $U(1)_{N}$ gauge symmetry is considerably
larger than in the NMSSM with three extra pairs of $5+\overline{5}$ supermultiplets of matter because of the
$U(1)_{N}$ $D$--term contribution to $m_{h_1}$ that increases the two--loop theoretical restriction on $m_{h_1}$
by $\sim 7-8\,\mbox{GeV}$. Thus this relatively small contribution to the lightest Higgs mass plays an important role enabling
us to find phenomenologically acceptable solutions with $125-126\,\mbox{GeV}$ Higgs mass near the quasi--fixed point (A)
in the case of scenario A.

\section{Conclusions}
In this paper we have explored the RG flow of the gauge and Yukawa couplings within the $E_6$ inspired SUSY models
with extra $U(1)_{N}$ gauge symmetry under which right--handed neutrinos have zero charge. In these models
single discrete $\tilde{Z}^{H}_2$ symmetry forbids the tree--level flavor--changing transitions and the most dangerous
baryon and lepton number violating operators.  Just below the GUT scale the matter content of these SUSY models includes
three copies of $27_i$--plets and a set of $M_{l}$ and $\overline{M}_l$ supermultiplets from the incomplete $27'_l$  and
$\overline{27'}_l$ representations of $E_6$. All supermultiplets $M_{l}$ are even under the $\tilde{Z}^{H}_2$ symmetry
whereas all matter superfields, that fill in complete $27_i$--plets, are odd. The supermultiplets $\overline{M}_l$ can be
either odd or even under the $\tilde{Z}^{H}_2$ symmetry. In particular, the set of the supermultiplets $M_{l}$ include
either lepton $SU(2)_W$ doublet $L_4$ (scenario A)  or colour triplet down type quark $d^c_4$ (scenario B) states to render
the lightest exotic quark unstable. In the scenario A the exotic quarks are leptoquarks while scenario B implies that the exotic
quarks are diquarks.

Our numerical analysis revealed that the solutions of the two--loop RG equations for the $SU(2)_W$ and $U(1)_Y$
gauge couplings are focused in the infrared region near the quasi--fixed points which are rather close to the measured values
of these couplings at the EW scale. On the other hand the convergence of the solutions for the strong gauge coupling $g_3(Q)$
to the fixed point is rather weak because the corresponding  one--loop beta function vanishes in the scenario A and remains quite
small, i.e. $\beta_3=1$, in the scenario B. Nonetheless we demonstrated that in the case of scenario A the values of the overall
gauge coupling $g_0$ around $1.5$ leads to $g_i(M_Z)$ which are quite close to the measured central values of
these couplings at the EW scale including the strong gauge coupling.  In the scenario B the low energy values of $g_3(Q)$
are always substantially smaller than the experimentally measured central value of this coupling. It means that the values of
$\alpha_3(M_Z)$, which are reasonably close to its measured value, result in the appearance of the Landau pole below the
GUT scale spoiling the gauge coupling unification in this scenario.

Moreover our analysis
indicates that in this case the SM--like Higgs state tends to be lighter than $125\,\mbox{GeV}$. Indeed, we argued that the
solutions of the two--loop RG equations for the Yukawa couplings are concentrated near the quasi--fixed points when $h_{t}(M_X)$
and $\lambda(M_X)$ grow. In the scenarios A and B these quasi--fixed points correspond to the values
of $\tan\beta$ around $1$ and $1.2$ respectively. Near the quasi--fixed point the low energy values of the coupling $\lambda$
tend to be slightly larger in the scenario A than in the scenario B.  As a consequence in the vicinity of the quasi--fixed point
the lightest Higgs state is allowed to be a few GeV heavier in the scenario A than in the scenario B. Our estimations show that
for $1.5\lesssim h_t(M_X),\,\lambda(M_X)\lesssim 3$ the maximal value of the lightest Higgs mass is just above $126\,\mbox{GeV}$
in the scenario A and a few GeV lower than $125\,\mbox{GeV}$ in the scenario B. Thus it seems to be rather problematic to find
phenomenologically acceptable solutions near the quasi--fixed point in the case of scenario B.

In this context it is worth noting that the absolute maximum value of the lightest Higgs mass in the $E_6$ inspired SUSY models
with extra $U(1)_{N}$ symmetry is about $155\,\mbox{GeV}$ \cite{King:2005jy} so that it is considerably larger than the upper
bounds on $m_{h_1}$ in the vicinity of the quasi--fixed points. This absolute maximum value of $m_{h_1}$ is
attained for $\tan\beta\simeq 1.5-2$ that correspond to the substantially lower values of $h_t(M_Z)$ than the ones associated
with the quasi--fixed point scenarios. Since larger values of the top--quark Yukawa coupling result in smaller $\lambda(M_Z)$
the two--loop upper bounds on $m_{h_1}$ near the quasi--fixed points are significantly lower than $155\,\mbox{GeV}$.
Besides the solutions with $125-126\,\mbox{GeV}$ SM--like Higgs boson can be obtained only if $\lambda(M_Z) > g'_1(M_Z)$.
Such solutions can be found in the vicinity of the quasi--fixed point in the case of scenario A. However for $\lambda(M_Z) > g'_1(M_Z)$
the Higgs spectrum has very hierarchical structure which implies that all Higgs states except the lightest one can not be discovered
at the LHC and the phenomenologically viable solutions associated with the quasi--fixed point scenarios are very fine--tuned.

Finally we compared the results of our analysis of the quasi--fixed point scenarios in the $E_6$ inspired SUSY models with the
corresponding results in the NMSSM and NMSSM+. The two--loop RG flow of the Yukawa couplings within the NMSSM+ is
very similar to the one in the scenario A. The low energy values of the Yukawa couplings and $\tan\beta$ associated with
the quasi--fixed point scenarios are very close in both models as well. Nevertheless because of the $U(1)_{N}$ $D$--term
contribution to $m_{h_1}$ the upper bound on the lightest Higgs boson mass is larger in the scenario A as compared with the
NMSSM+. As a result in the NMSSM+ the theoretical restriction on $m_{h_1}$ in the vicinity of the quasi--fixed point is
lower than $120\,\mbox{GeV}$. This does not rule out NMSSM+ but definitely disfavours the corresponding scenario.
In the NMSSM the top--quark Yukawa coupling is smaller whereas $\lambda(M_Z)$ and $\tan\beta$ are larger near the
quasi--fixed point as compared with the quasi--fixed point scenarios in the $E_6$ inspired SUSY models discussed here
and NMSSM+. Therefore the upper bound on the mass of the lightest Higgs boson is less stringent and can be almost
as large as $130\,\mbox{GeV}$.


The results presented in this article show that it is not so easy to get $125-126\,\mbox{GeV}$ SM--like Higgs
mass within the non--minimal SUSY models mentioned above as one could naively expect. In this context it would be
appropriate to remind that in the MSSM large loop corrections are required to raise the Higgs boson mass to
$125\,\mbox{GeV}$. This can be achieved only if stops are relatively heavy that leads to some degree of fine-tuning.
As follows from Tables 4 and 5 near the quasi--fixed points in the case of Scenario A and NMSSM the tree-level upper
bound on the lightest Higgs boson mass may be still 10 GeV larger than in the MSSM so that in order to match the
$125-126\,\mbox{GeV}$ Higgs mass value the size of loop corrections can be smaller and stops can be lighter in these
cases as compared with the ones in the minimal SUSY model.

\section*{Acknowledgements}
\vspace{-2mm} We would like to thank P.~Athron, A.~V.~Borisov, S.~F.~King, S.~Pakvasa, O.~Pavlovsky, A.~Thomas, X.~Tata,
M.~Trusov and A.~Williams for  fruitful discussions. R.N. is also grateful to E.~Boos, M.~Dubinin, S.~Demidov, D.~Gorbunov,
M.~Libanov, V.~Novikov, O.~Kancheli, D.~Kazakov, V.~Rubakov, M.~Shifman, S.~Troitsky, M.~Vysotsky for valuable comments
and remarks. This work was supported by the University of Adelaide and the Australian Research Council through the
ARC Centre of Excellence for Particle Physics at the Terascale (CoEPP).

\newpage
\begin{table}[tab2]
\centering
\begin{tabular}{|c|c|c|c|c|c|c|c|c|}
\hline
                    & $g_0$  & $g_3$         &  $g_2$          & $g_1$           & $g'_1$           & $g_{11}$           & $\ds\frac{g^2_1}{g^2_2}$ & $\ds\frac{g'_1}{g_1}$\\
\hline
                    & 1.2       &    1.074       &   0.628          &  0.454          &  0.458            &    0.0196            &   0.523   &   1.0090      \\
\cline{2-9}
                    & 1.5       &    1.213       &   0.655          &  0.465          &  0.469            &    0.0210            &   0.503   &   1.0090      \\
Scenario A   &             &    ( 1.5 )       & (0.684)         & (0.471)        &  (0.476)          &   (0.0219)           &  (0.474) &  (1.0106)    \\
\cline{2-9}
                    & 2.0       &    1.330        &  0.676          &  0.473           &  0.477            &    0.0221            &   0.489   &   1.0084      \\
\cline{2-9}
                    & 3.0       &    1.395        &  0.689          &  0.478           &  0.481            &    0.0228            &   0.481   &   1.0074     \\
                    &             &    ( 3.0 )        & (0.744)        & (0.489)          & (0.495)          &   (0.0246)           &  (0.432)  &  (1.0116)  \\
\hline
                    & 1.2       &    0.881        &  0.582          &  0.436           &  0.431            &    -0.0254           &   0.560   &   0.989      \\
\cline{2-9}
                    & 1.6       &    0.975        &  0.609          &  0.447           &  0.442            &    -0.0275           &   0.539   &   0.988      \\
Scenario B   &             &   (1.108)      & (0.632)         & (0.453)         & (0.448)           &   (-0.0286)         &  (0.514)  &  (0.989)   \\
\cline{2-9}
                    & 2.0       &    1.020        &  0.622          &  0.453            & 0.447             &   -0.0285            &  0.530    &   0.987      \\
\cline{2-9}
                    & 3.0       &    1.057        &  0.633          &  0.458            & 0.451             &   -0.0294            & 0.523     &   0.986    \\
                    &             &   (1.368)      &  (0.670)        & (0.466)          & (0.461)           &  (-0.0312)          & (0.484)   &   (0.989)  \\
\hline
\end{tabular}
\caption{The values of the gauge couplings at the EW scale. These couplings are calculated
for $g_1(M_X)=g'_1(M_X)=g_2(M_X)=g_3(M_X)=h_t(M_X)=\lambda(M_X)=g_0$,
$g_{11}(M_X)=0$ and different values of $g_0$ in the two--loop approximation.
The low--energy values of the corresponding couplings calculated in the one--loop approximation
are given in the brackets.}
  \label{tab2}
\end{table}

\vspace{3cm}
\begin{table}[tab3]
\centering
\begin{tabular}{|c|c|c|c|c|c|c|c|c|}
\hline
                   & $g_0$ & $h_t(M_X)$ & $\lambda(M_X)$ & $h_t(M_Z)$ & $\lambda(M_Z)$  & $\tan\beta$ & $m^{(0)}_{h_1}$ & $m^{(2)}_{h_1}$ \\
                  &             &                    &                            &                     &                             &                    &  (GeV)              &  (GeV)      \\
\hline
                   &  1.5     &    3.0           &     3.0                  &  1.31            &    0.46                  &    1.02         &   92.2               &  120.5    \\
\cline{2-9}
Scenario A  &  1.5     &    1.5           &     3.0                  &  1.30            &    0.53                  &    1.05         &   102.4             &  126.6      \\
\cline{2-9}
                   &   1.5     &    3.0           &     1.5                  &  1.32            &   0.37                  &    1.01         &   79.0               &  113.3     \\
\cline{2-9}
                   &   1.5     &    1.5           &     1.5                  &  1.31            &   0.46                  &    1.03         &   91.7               &  120.3     \\
\hline
                   &   3.0     &    3.0           &     3.0                  &  1.22            &   0.44                  &    1.20         &   88.7               &  119.1  \\
\cline{2-9}
Scenario B  &   3.0     &    1.5           &     3.0                  &  1.21             &   0.49                 &    1.22         &   95.7               &  123.2     \\
\cline{2-9}
                   &   3.0     &    3.0           &     1.5                  &  1.23             &   0.36                  &    1.19         &   76.6               &  112.3    \\
\cline{2-9}
                   &   3.0     &    1.5           &     1.5                  &  1.22             &   0.43                  &    1.20         &   86.0               &  117.5     \\
\hline
\end{tabular}
\caption{The values of the Yukawa couplings at the EW scale and the upper bounds on the lightest
Higgs mass in the scenarios A and B. The values of $h_t(M_Z)$ and $\lambda(M_Z)$ are calculated using
two--loop RG equations for $g_1(M_X)=g'_1(M_X)=g_2(M_X)=g_3(M_X)=g_0$, $g_{11}(M_X)=0$ and
different values of $h_t(M_X)$ and $\lambda(M_X)$. The low energy values of the Yukawa couplings are used
for the calculation of $\tan\beta$, tree--level and two--loop upper bounds on the mass of the lightest Higgs
boson ($m^{(0)}_{h_1}$ and $m^{(2)}_{h_1}$ respectively). We set $m_t(M_t)=163\,\mbox{GeV}$,
$M_S=1200\,\mbox{GeV}$ and $X_t=\sqrt{6}M_S$.}
\label{tab3}
\end{table}

\newpage
\begin{table}[tab4]
\centering
\begin{tabular}{|c|c|c|c|c|c|c|c|c|}
\hline
                   & $g_0$ & $h_t(M_X)$ & $\lambda(M_X)$ & $h_t(M_Z)$ & $\lambda(M_Z)$  & $\tan\beta$ & $m^{(0)}_{h_1}$ & $m^{(2)}_{h_1}$ \\
                  &             &                    &                            &                     &                             &                    &  (GeV)              &  (GeV)      \\
\hline
                   &  0.725 &    3.0           &     3.0                  &  1.10            &    0.54                  &    1.60         &   93.4               &  122.5    \\
\cline{2-9}
NMSSM      &  0.725 &    1.5           &     3.0                  &  1.06             &    0.62                 &    1.90         &   102.8             &  128.9      \\
\cline{2-9}
                   &  0.725 &    3.0           &     1.5                  &  1.12             &   0.43                  &    1.52         &   77.8               &  113.2     \\
\cline{2-9}
                   &  0.725 &    1.5           &     1.5                  &  1.08             &   0.53                  &    1.73         &   92.2               &  121.9     \\
\hline
                   &   1.5    &    3.0           &     3.0                  &  1.31             &   0.45                  &    1.02         &   78.3               &  113.0   \\
\cline{2-9}
NMSSM+    &   1.5    &    1.5           &     3.0                  &  1.29             &   0.51                  &    1.05         &   88.5               &  118.5      \\
\cline{2-9}
                   &   1.5     &    3.0           &     1.5                  &  1.32             &   0.35                 &    1.01          &   61.3               &  105.0    \\
\cline{2-9}
                   &   1.5     &    1.5           &     1.5                  &  1.30             &   0.43                 &    1.03          &   74.2               &  111.0     \\
\hline
\end{tabular}
\caption{The values of the Yukawa couplings at the EW scale and the upper bounds on the lightest
Higgs mass in the NMSSM and NMSSM+. The values of $h_t(M_Z)$ and $\lambda(M_Z)$ are calculated using
two--loop RG equations for $g_1(M_X)=g_2(M_X)=g_3(M_X)=g_0$ and different values of $h_t(M_X)$ and
$\lambda(M_X)$. The low energy values of the Yukawa couplings are used for the calculation of $\tan\beta$,
tree--level and two--loop upper bounds on the mass of the lightest Higgs boson ($m^{(0)}_{h_1}$ and
$m^{(2)}_{h_1}$ respectively). We set $m_t(M_t)=163\,\mbox{GeV}$, $M_S=1200\,\mbox{GeV}$ and
$X_t=\sqrt{6}M_S$.}
\label{tab4}
\end{table}

\newpage
\begin{figure}
\includegraphics[width=75mm,height=75mm]{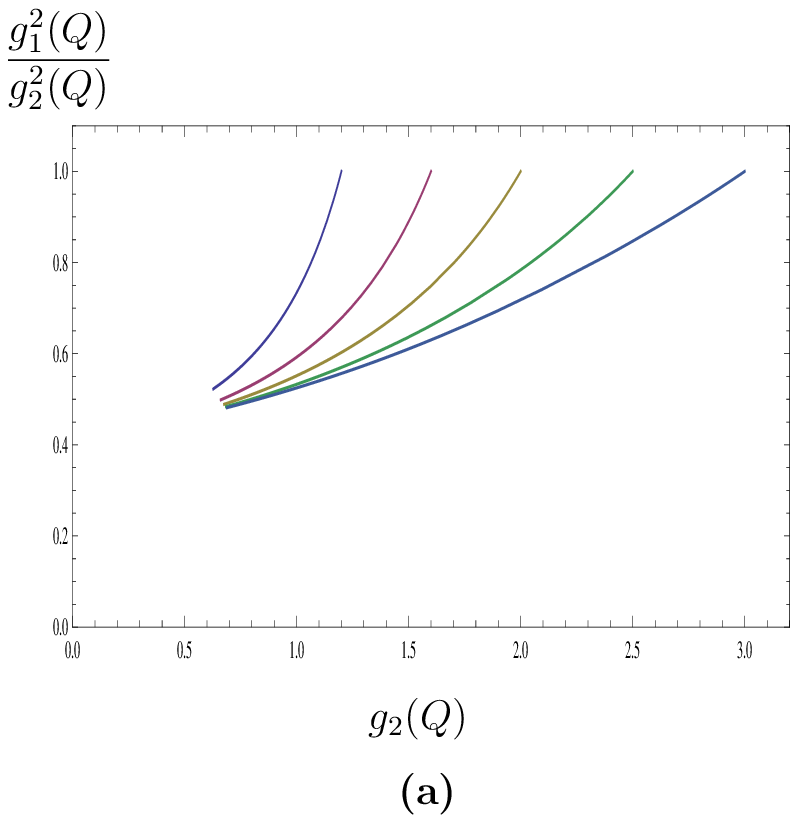}\qquad
\includegraphics[width=75mm,height=75mm]{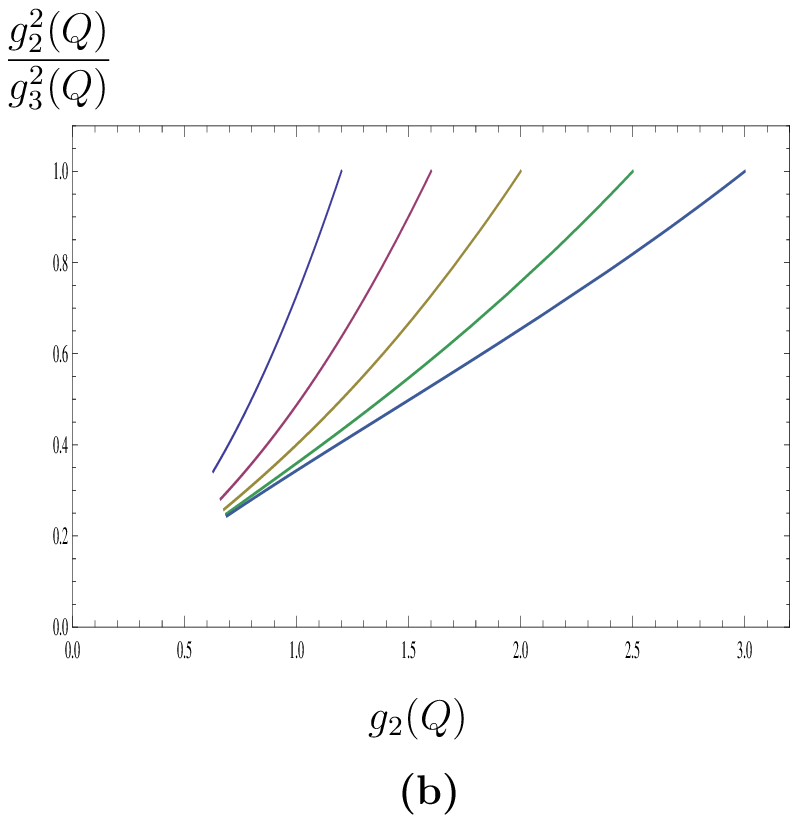}\\
\includegraphics[width=75mm,height=75mm]{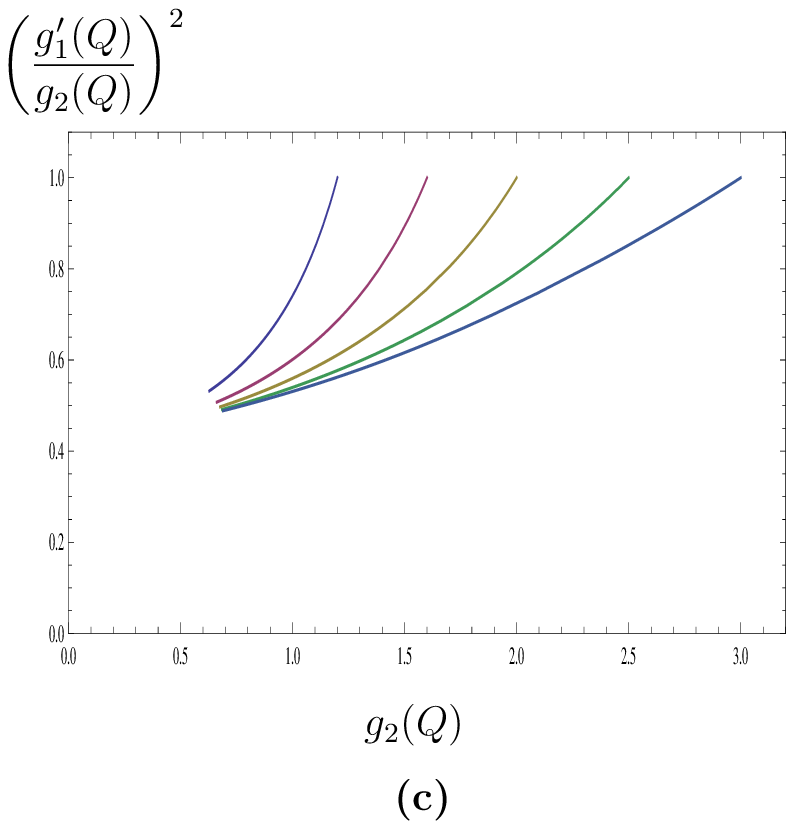}\qquad
\includegraphics[width=75mm,height=75mm]{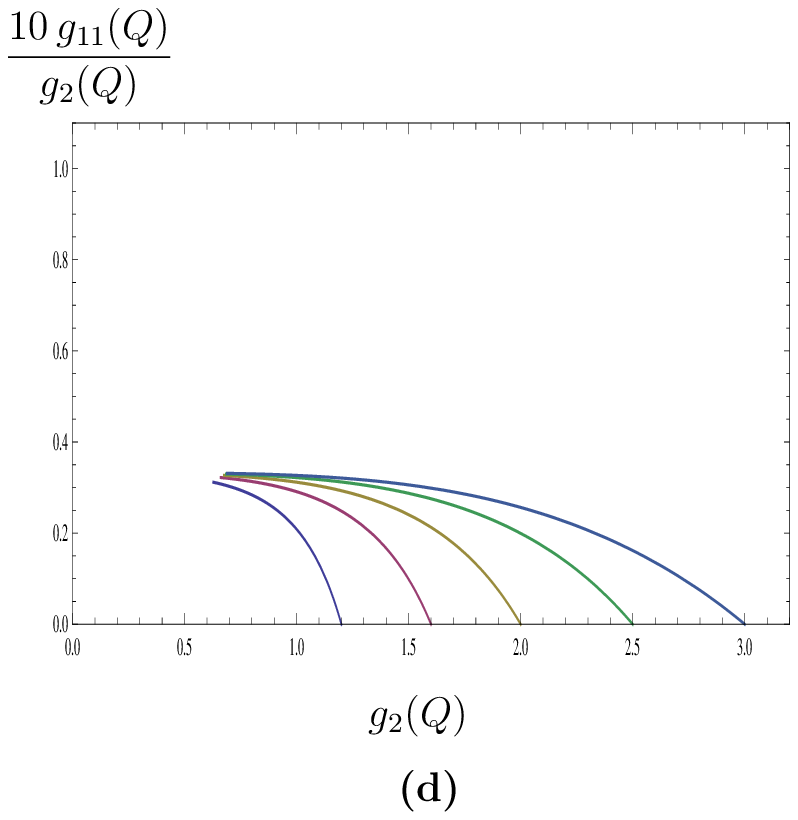}\\
\caption{Two--loop RG flow of gauge couplings in the case of scenario A
for $g_1(M_X)=g'_1(M_X)=g_2(M_X)=g_3(M_X)=h_t(M_X)=\lambda(M_X)=g_0$,
$g_{11}(M_X)=0$ and different values of $g_0$: {\it (a)} evolution of
$\ds\frac{g^2_1(Q)}{g^2_2(Q)}$ versus $g_2(Q)$ from $Q=M_X$ to the EW scale;
{\it (b)} running of $\ds\frac{g_2^2(Q)}{g_3^2(Q)}$ versus $g_2(Q)$ from
$Q=M_X$ to the EW scale; {\it (c)} RG flow of $\ds\frac{g^{'2}_1(Q)}{g^2_2(Q)}$
versus $g_2(Q)$ from $Q=M_X$ to the EW scale; {\it (d)} running of
$\ds \frac{10\, g_{11}(Q)}{g_2(Q)}$ versus $g_2(Q)$ from $Q=M_X$ to the EW scale.
}
\label{essmfig1}
\end{figure}

\newpage
\begin{figure}
\includegraphics[width=75mm,height=75mm]{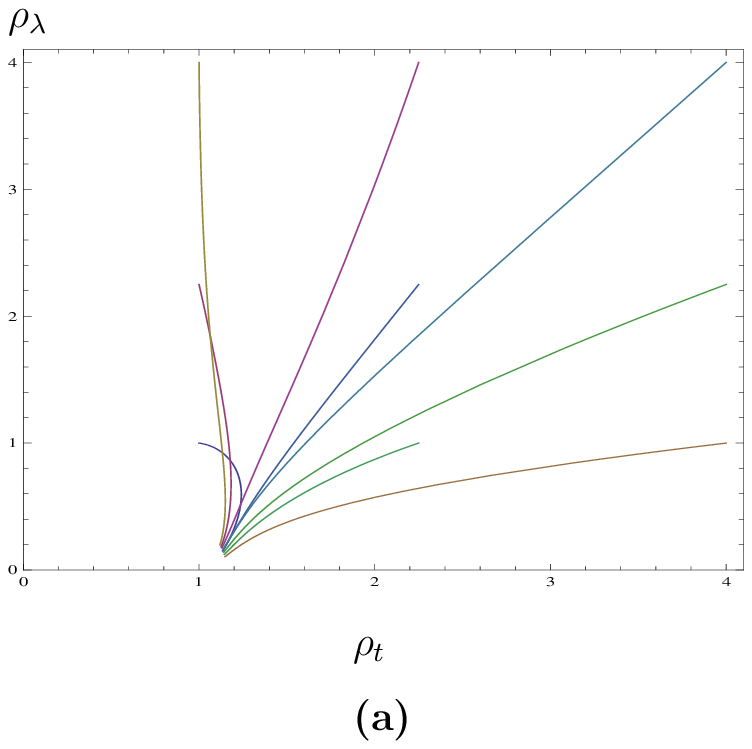}\qquad
\includegraphics[width=75mm,height=75mm]{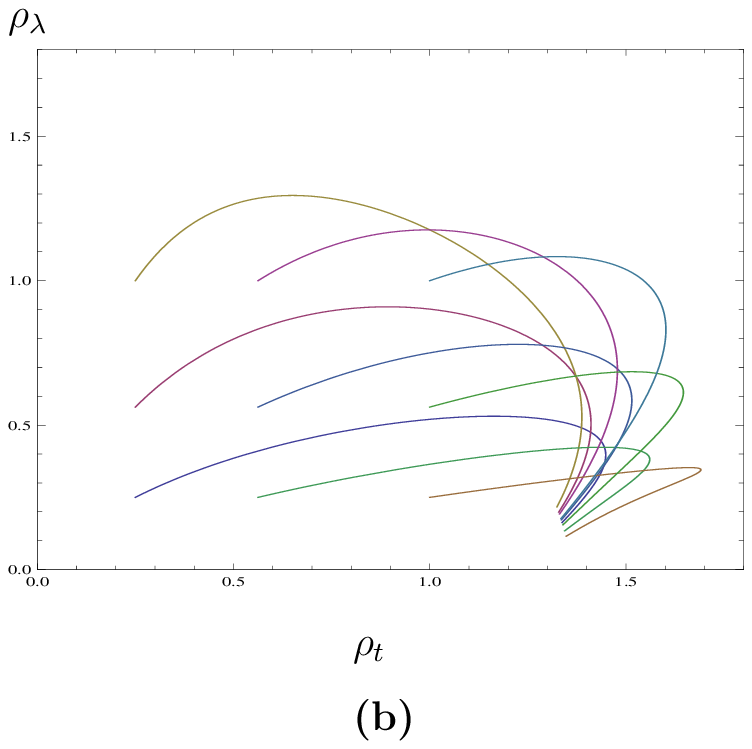}\\
\includegraphics[width=75mm,height=75mm]{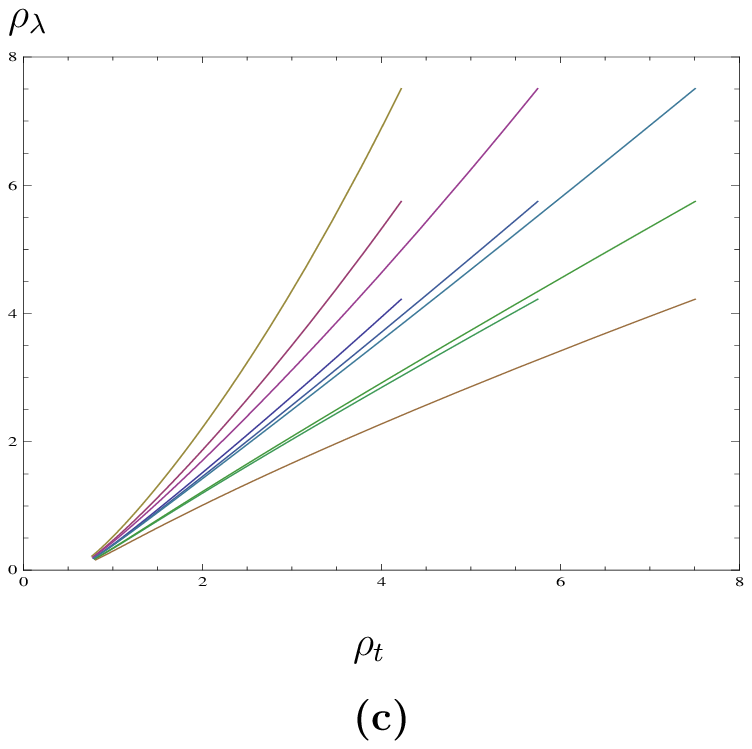}\qquad
\includegraphics[width=75mm,height=75mm]{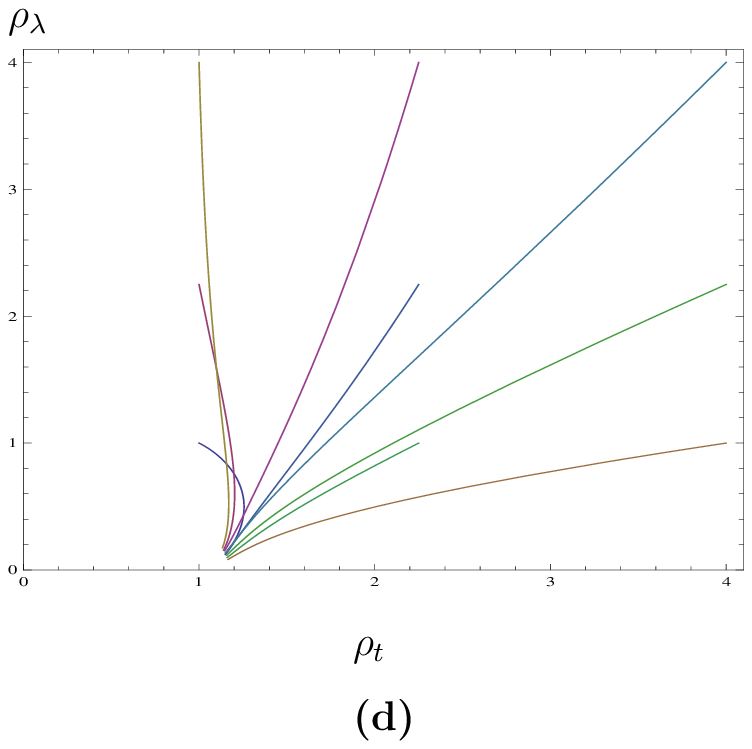}\\
\caption{(a) Two--loop RG flow of $\rho_{\lambda}$ versus $\rho_{t}$ in
the scenario A for $g_0=1.5$. (b) Two--loop RG flow of the Yukawa couplings in the
$\rho_{\lambda}-\rho_{t}$ plane in the scenario B for $g_0=3$. (c) Two--loop RG flow of
$\rho_{\lambda}$ versus $\rho_{t}$ within the NMSSM for $g_0=0.725$. (d) Two--loop RG
flow of the Yukawa couplings in the $\rho_{\lambda}-\rho_{t}$ plane within the NMSSM+
for $g_0=1.5$.  In all cases the energy scale $Q$ is varied from $M_X$ to $M_Z$.
Different trajectories correspond to different initial conditions for $\lambda$ and $h_{t}$
at the scale $M_X$.}
\label{essmfig2}
\end{figure}

\end{document}